\documentclass[journal=jacsat,manuscript=article]{achemso}

\usepackage[version=3]{mhchem} 
\usepackage{xcolor,isomath,upgreek,comment,braket,amsmath,amssymb,hyperref,cleveref,textcomp}

\author{Tianqi Zhu}
\affiliation{Department of Physics, ETH Z{\"u}rich, Otto-Stern-Weg 1, 8093 Z{\"u}rich, Switzerland}
\author{Jan Rhensius}
\affiliation{QZabre LLC, Regina-Kägi-Strasse 11, 8050 Z{\"u}rich, Switzerland}
\author{Viraj Damle}
\affiliation{QZabre LLC, Regina-Kägi-Strasse 11, 8050 Z{\"u}rich, Switzerland}
\author{Konstantin Herb}
\affiliation{Department of Physics, ETH Z{\"u}rich, Otto-Stern-Weg 1, 8093 Z{\"u}rich, Switzerland}
\author{Gabriel Puebla-Hellmann}
\affiliation{QZabre LLC, Regina-Kägi-Strasse 11, 8050 Z{\"u}rich, Switzerland}
\author{Christian L. Degen}
\affiliation{Department of Physics, ETH Z{\"u}rich, Otto-Stern-Weg 1, 8093 Z{\"u}rich, Switzerland}
\email{degenc@ethz.ch}
\author{Erika Janitz}
\affiliation{Department of Physics, ETH Z{\"u}rich, Otto-Stern-Weg 1, 8093 Z{\"u}rich, Switzerland}
\email{ejanitz@phys.ethz.ch}

\title[An \textsf{achemso} demo]
  {Multicone Diamond Waveguides for Nanoscale Quantum Sensing}


\begin{document}


\begin{abstract}
The long-lived electronic spin of the nitrogen-vacancy (NV) center in diamond is a promising quantum sensor for detecting nanoscopic magnetic and electric fields in a variety of experimental conditions. Nevertheless, an outstanding challenge in improving measurement sensitivity is the poor signal-to-noise ratio (SNR) of prevalent optical spin-readout techniques. Here, we address this limitation by coupling individual NV centers to optimized diamond nanopillar structures, thereby improving optical collection efficiency of fluorescence. First, we optimize the structure in simulation, observing an increase in collection efficiency for tall ($\geq5 \ \upmu$m) pillars with tapered sidewalls. We subsequently verify these predictions by fabricating and characterizing a representative set of structures using a reliable and reproducible nanofabrication process. An optimized device yields increased SNR, owing to improvements in collimation and directionality of emission. Promisingly, these devices are compatible with low-numerical-aperture, long-working-distance collection optics, as well as reduced tip radius, facilitating improved spatial resolution for scanning applications. 
\end{abstract}

\section{Introduction}
Electronic spins associated with individual atomic defects in wide-bandgap materials~\cite{aharonovich2016solid} can serve as magnetic sensors with exquisite sensitivity and nanoscale spatial resolution\cite{schirhagl_arpc_2014}. Most notably, near-surface nitrogen-vacancy (NV) centers in diamond\cite{dohertyNitrogenvacancyColourCentre2013} have been harnessed to image exotic magnetic materials,~\cite{gross2017real,thiel_science_2019} nanoscale currents,~\cite{chang_nanolett_2017} and single- to few-molecule samples\cite{lovchinsky2016nuclear}. Such experiments profit from the exceptional spin coherence of the NV, which can exceed $1$ ms at room temperature~\cite{herbschleb_natcomm_2019} 
and $1$ s at cryogenic temperatures.\cite{abobeih2018one} In addition, the spin state can be efficiently initialized with a laser\cite{robledo2011spin} and manipulated with microwave fields.\cite{rong2015experimental} Despite their promise, measurement sensitivities for near-surface NVs are hampered by poor signal-to-noise ratios (SNRs) for optical spin readout.\cite{taylorHighsensitivityDiamondMagnetometer2008} Indeed, $\mathrm{SNR}$ scales with the square root of the collected fluorescence, which is limited by the high refractive index of diamond ($n_d=2.4$), causing total internal reflection for emission outside of a critical angle of $\approx25^\circ$\cite{castellettoDiamondbasedStructuresCollect2011}. Consequently, improved collimation of NV-center emission is highly desirable.  

Promisingly, recent progress in nanophotonics~\cite{castelletto2017advances} aims to obviate this challenge by coupling fluorescence to optical structures with improved collection efficiency.~\cite{schroder2016quantum} This progress can be divided into two categories: 1) hybrid nanophotonics, including structures fabricated from alternative materials that are interfaced with diamond, and 2) diamond nanophotonics, where optical devices are carved into the diamond itself. Hybrid approaches may benefit from straightforward and mature fabrication techniques available for {\it e.g.}, metals,\cite{choyEnhancedSinglephotonEmission2011} silica,\cite{parkCavityQEDDiamond2006} Si,\cite{chakravarthiInversedesignedPhotonExtractors2020} and III-V materials.\cite{englundDeterministicCouplingSingle2010,wamboldAdjointoptimizedNanoscaleLight2021} However, such structures often exhibit reduced coupling to NV emission~\cite{lonvcar2013quantum}, and the presence of additional materials may preclude nanoscale proximity to sensing targets.~\cite{baiHybridDiamondGlassOptical2018} 

In contrast, diamond optical structures facilitate maximum coupling to NV centers and proximity to sensing targets. Moreover, despite the mechanical resilience of diamond, recent nanofabrication advances~\cite{burek2014high,mitchell2019realizing,atikian2017freestanding}  
have paved the way for creating bespoke diamond devices such as lenses \cite{haddenStronglyEnhancedPhoton2010,huangMonolithicImmersionMetalens2019}, gratings\cite{liEfficientPhotonCollection2015}, optical cavities\cite{burek2014high,mitchell2019realizing,hausmann2013coupling}, and waveguides\cite{hausmann2010fabrication,babinec_natnano_2010}. 
In particular, nanoscale pillars containing shallow NVs 
have garnered attention for sensing applications. These devices act as waveguides for NV fluorescence, offering broadband enhancement in collection efficiency. 
In addition, $\sim100$ nm device diameters facilitate exquisite ($<50$ nm) spatial resolution for scanning NV applications,~\cite{marchiori2022nanoscale} which utilize the diamond pillar as a probe for atomic-force microscopy (AFM).~\cite{degenScanningMagneticField2008,maletinskyRobustScanningDiamond2012}
Already, several generations of devices have been explored in the literature, including cylinders,\cite{babinec_natnano_2010} truncated cones,\cite{momenzadehNanoengineeredDiamondWaveguide2015,torun2021optimized} and parabolic reflectors.\cite{wanEfficientExtractionLight2018,hedrichParabolicDiamondScanning2020}. 
Despite this progress, further understanding and optimization of device geometry represents a key challenge toward improving optical spin-readout efficiency and thereby measurement sensitivity.

In this work, we explore the impact of nanopillar geometry on collection efficiency for NV-center fluorescence. Building on a truncated-cone design~\cite{momenzadehNanoengineeredDiamondWaveguide2015} (Fig. \ref{fig:1}a), we develop a simulation model that predicts improved fluorescence collimation for tall pillars. Moreover, we observe enhanced collimation and emission directionality by inclusion of a second, shallow sidewall angle near the pillar facet. Armed with these predictions, we develop a simple, reliable, and reproducible fabrication process for realizing a representative set of structures. Subsequent optical characterization yields a maximum spin-readout $\mathrm{SNR}=0.106$ for the optimized geometry, corresponding to an expected factor-of-three improvement compared to a cylindrical device of similar dimensions. Moving forward, the design and fabrication principles developed in this work will facilitate a new generation of efficient devices with superior sensitivity and exceptional spatial resolution. 

\section{Simulations} \label{sec:sims}
\begin{figure}
\includegraphics{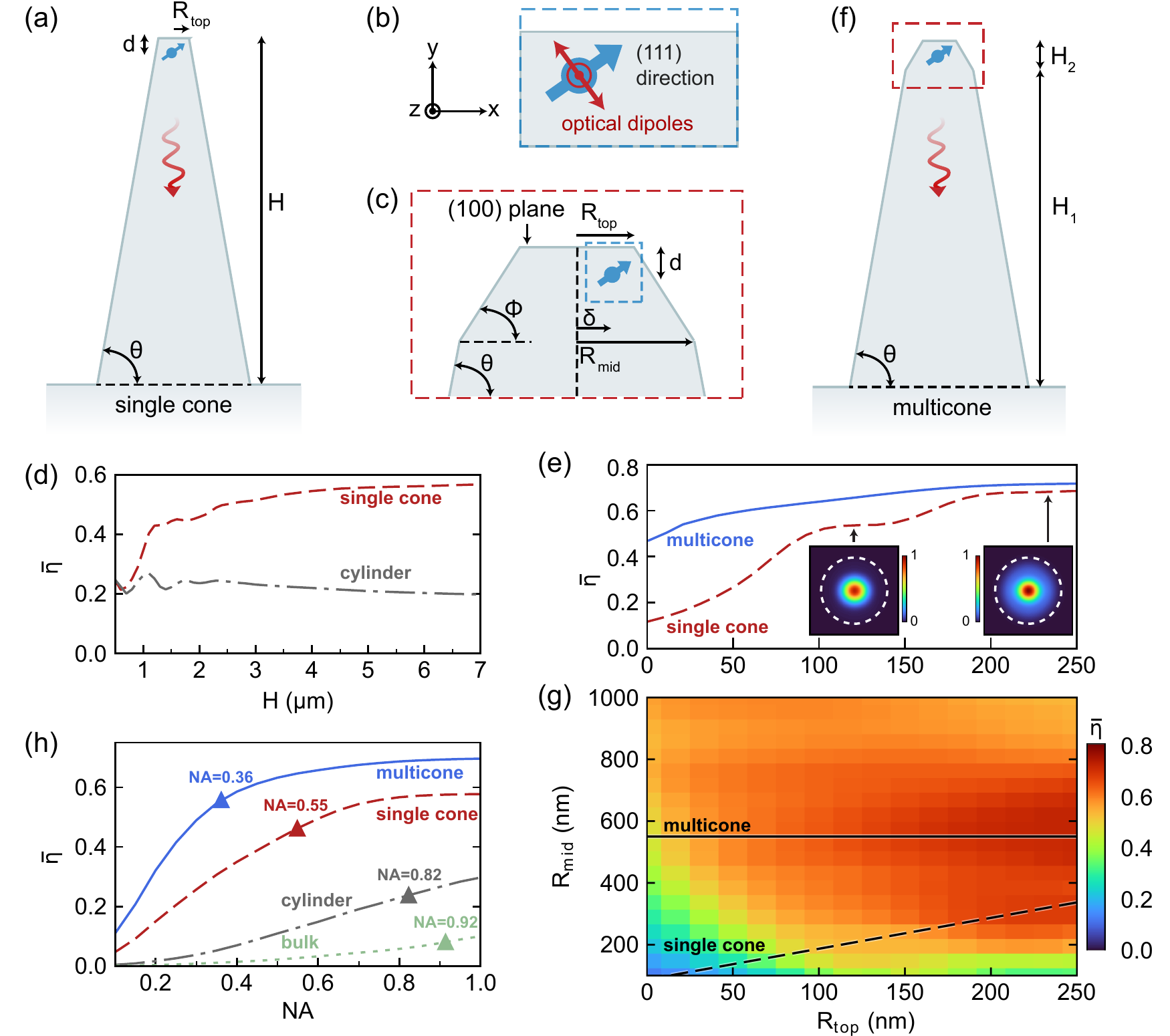}
   \caption{
   Nanopillar design and simulations. (a) Schematic of the single-cone geometry. (b) Illustration of an NV center oriented along the [111] crystal axis. (c) Example pillar facet including two distinct sidewall angles ($\theta,\phi$) {\it i.e.,} the multicone geometry. (d) Collection efficiency $\bar{\eta}$ as a function of $H$ for single-cone and cylinder geometries ($R_\mathrm{{top}}=150$ nm for both). (e) $\bar{\eta}$ as a function of $R_{\mathrm{top}}$ for the single-cone (dashed) and multicone (solid) geometries ($H=5\ \upmu$m for both). Far-field modal distributions (in air) are plotted for the single cone for $R_{\mathrm{top}}= 123$ and 238 nm. Circles corresponding to the experimental $\mathrm{NA}=0.75$ are overlaid. (f) Schematic of the multicone geometry. (g) $\bar{\eta}$ as a function of $R_{\mathrm{top}}$ and $R_{\mathrm{mid}}$ for a multicone device. We consider two sidewall angles (fixed $\theta=80^\circ$ and varying $\phi$) with $H_\mathrm{1}=4.5$ $\upmu$m and $H_\mathrm{2}=0.5$ $\upmu$m, respectively. Lines corresponding to the single-cone and multicone structures in (e) are overlaid. (h) $\bar{\eta}$ {\it vs.} NA for three pillar geometries ($R_\mathrm{{top}}=150$ nm, $H=5$ $\upmu$m) and bulk diamond. Triangles indicate $\mathrm{NA}_{\,0.80}$; the NA at which 80\% of the intensity for $\mathrm{NA}=1$ is collected. 
   \label{fig:1}}
\end{figure}
We simulate the optical properties of diamond nanopillars using a finite-difference time-domain software from Lumerical Inc.~\cite{noauthor_photonics_nodate}. As a figure of merit, we calculate the wavelength-dependent collection efficiency $\eta\left(\lambda\right)$ normalized to the NV emission spectrum at room temperature $I\left(\lambda\right)$,~\cite{rondinSurfaceinducedChargeState2010} yielding  
\begin{align}
\bar{\eta} = \frac{\int_{\lambda_{\mathrm{coll}}} \eta(\lambda) I(\lambda )d\lambda}{\int_{\lambda_{\mathrm{coll}}} {I(\lambda )} d\lambda}.
\end{align}
Here, we consider free-space optical collection from the base of the pillar with numerical aperture $\mathrm{NA}=0.75$ and bandwidth $\lambda_{\mathrm{coll}}=650-800$ nm (matching our experimental setup). Individual NV centers are modeled as two orthogonal electric dipoles located in the plane perpendicular to the [111] direction of a (100)-cut diamond (Fig. \ref{fig:1}b).~\cite{epsteinAnisotropicInteractionsSingle2005} Moreover, we assume all defects are located $d=5$ nm below the pillar facet and centered laterally within the structure ($\delta=0$, Fig. \ref{fig:1}c) as we observe little variation in $\bar{\eta}$ over the range of possible emitter positions (see Supporting Information or SI, Fig. S2).

Our starting point for developing an optimized nanopillar is the truncated-cone geometry (hereafter referred to as the ``single cone" or SC), \cite{momenzadehNanoengineeredDiamondWaveguide2015} which is fully parameterized by the top radius $R_{\mathrm{top}}$, height $H$, and sidewall angle $\theta$ (Fig.~\ref{fig:1}a). First, we explore the impact of $H$ on $\bar{\eta}$ for $R_{\mathrm{top}}=150$ nm and $\theta=80^{\circ}$ (Fig. \ref{fig:1}d). We limit our study to $H=1-7$ $\upmu$m, which is the range over which we can achieve high fabrication yield (larger $H$ and smaller $R_\mathrm{top}$ result in device breakage). 
Moreover, $\theta$ was chosen to match the results of our standard fabrication process, which yields $\theta=78-86^\circ$ (see Fabrication section). 

Interestingly, the SC exhibits increasing $\bar{\eta}$ {\it vs.} $H$ that saturates for $H\gtrsim5$ $\upmu$m, which can be attributed to adiabatic expansion of the beam as it propagates down the structure~\cite{munschDielectricGaAsAntenna2013,stepanovHighlyDirectiveGaussian2015,hedrichParabolicDiamondScanning2020}. Indeed, at the bulk-diamond interface, the waveguide has expanded by a factor of $\mathrm{\Delta}\approx1+H\cot\theta/R_{\mathrm{top}}$, reducing the divergence angle of the exiting beam by a similar factor (in the paraxial limit). In contrast, an equivalent cylindrical pillar ($\theta=90^\circ$, Fig.~\ref{fig:1}d) exhibits a slight decay in $\bar{\eta}$ for tall structures. Furthermore, the tapered sidewalls of the SC increase total internal reflection at the top of the device, yielding $\bar{\eta}>0.5$ for tall devices. Subsequently, we target $H=5\ \upmu$m to simultaneously achieve high collection efficiency and fabrication yield.

Next, we simulate $\bar{\eta}$ as a function of $R_{\mathrm{top}}$ (Fig.~\ref{fig:1}e) and observe a monotonic relationship. Specifically, $\bar{\eta}$ increases rapidly for small $R_{\mathrm{top}}$ and plateaus at $\approx0.5$ for $R_{\mathrm{top}}=100-150$ nm, corresponding to the radii at which the waveguide supports fundamental modes within the NV spectrum. Collection efficiency continues to increase for $R_{\mathrm{top}}>150$ nm, with further plateaus corresponding to support of higher-order transverse modes. However, it is simultaneously desirable to reduce $R_{\mathrm{top}}$ to maximize spatial resolution for AFM applications. Consequently, we target $R_{\mathrm{top}}=150$ nm as a compromise between high $\bar{\eta}$, device yield, and spatial resolution in scanning experiments. 

Recent results demonstrate that a parabolic structure can improve collection efficiency owing to enhanced reflection at the top of the pillar.\cite{wanEfficientExtractionLight2018,hedrichParabolicDiamondScanning2020} Inspired by these results, we explore the impact of introducing an additional sidewall angle $\phi$ near the facet (Figs. \ref{fig:1}c and f), yielding a ``multicone" (MC) structure. Again, we consider a $5$-$\upmu$m-tall device with a lower region ($H_1=4.5\ \upmu$m) defined by our standard fabrication procedure ($\theta=80^\circ$) and an upper region ($H_2=0.5\ \upmu$m) with varying sidewall angle $\phi$. Equivalently, the upper region can be parameterized by the radius at the interface between regions $R_{\mathrm{mid}}=R_{\mathrm{top}}+H_2\cot{\phi}$. We simulate $\bar{\eta}$ for varying $R_{\mathrm{top}}$ and $R_{\mathrm{mid}}$ (Fig. \ref{fig:1}g) and overlay a dashed line corresponding to the SC ($\phi=80^\circ$) for comparison; we observe that for every $R_{\mathrm{top}}$, a MC geometry with $\phi<\theta$ exists that yields larger $\bar{\eta}$. Moreover, the parameter space for improved collection efficiency is relatively large ($\bar{\eta}>0.6$ for $R_{\mathrm{top}}=130-250$ nm and $\phi=40-80^\circ$), providing generous fabrication tolerances. 

Similar to the SC, the improvement in $\bar{\eta}$ for the MC can be partially ascribed to the increased mode diameter at the base of the pillar. However, introduction of a shallow sidewall angle at the facet also increases reflection at this interface and thereby the directionality of emission. To gain further insight into the role of $R_{\mathrm{top}}$, we overlay a solid line in Fig. \ref{fig:1}g for $R_{\mathrm{mid}}=550$ nm and plot the corresponding $\bar{\eta}$ in Fig. \ref{fig:1}e. In contrast to the plateaus observed for the SC, the MC exhibits a relatively smooth increase in $\bar{\eta}$ {\it vs.} $R_{\mathrm{top}}$ that can be attributed to transverse-mode mixing caused by the reduced sidewall angle, which no longer fulfills the adiabatic expansion criteria.~\cite{milton_mode_1977,fu_efficient_2014} For free-space or multimode collection, the ability to support additional transverse modes further increases $\bar{\eta}$.

We summarize our pillar optimization by plotting $\bar{\eta}$ {\it vs.} NA for a cylinder, SC, MC (all with $R_\mathrm{top}=150$ nm and $H=5$ $\upmu$m), and bulk diamond (Fig. \ref{fig:1}h). As expected, the MC yields the highest collection efficiency for NA in the range [0.1,1]. In addition, we compare the collimation achieved by each structure by calculating $\mathrm{NA}_{\,0.80}$; the NA yielding 80\% of the collected intensity for $\mathrm{NA}=1$ (triangular markers in Fig. \ref{fig:1}h). Again, the MC exhibits the best performance with $\mathrm{NA}_{\,0.8}=0.36$. This compatibility with low-NA {\it i.e.,} long-working-distance collection optics will reduce experimental costs and complexity in subsequent applications. 

\begin{figure}
\includegraphics{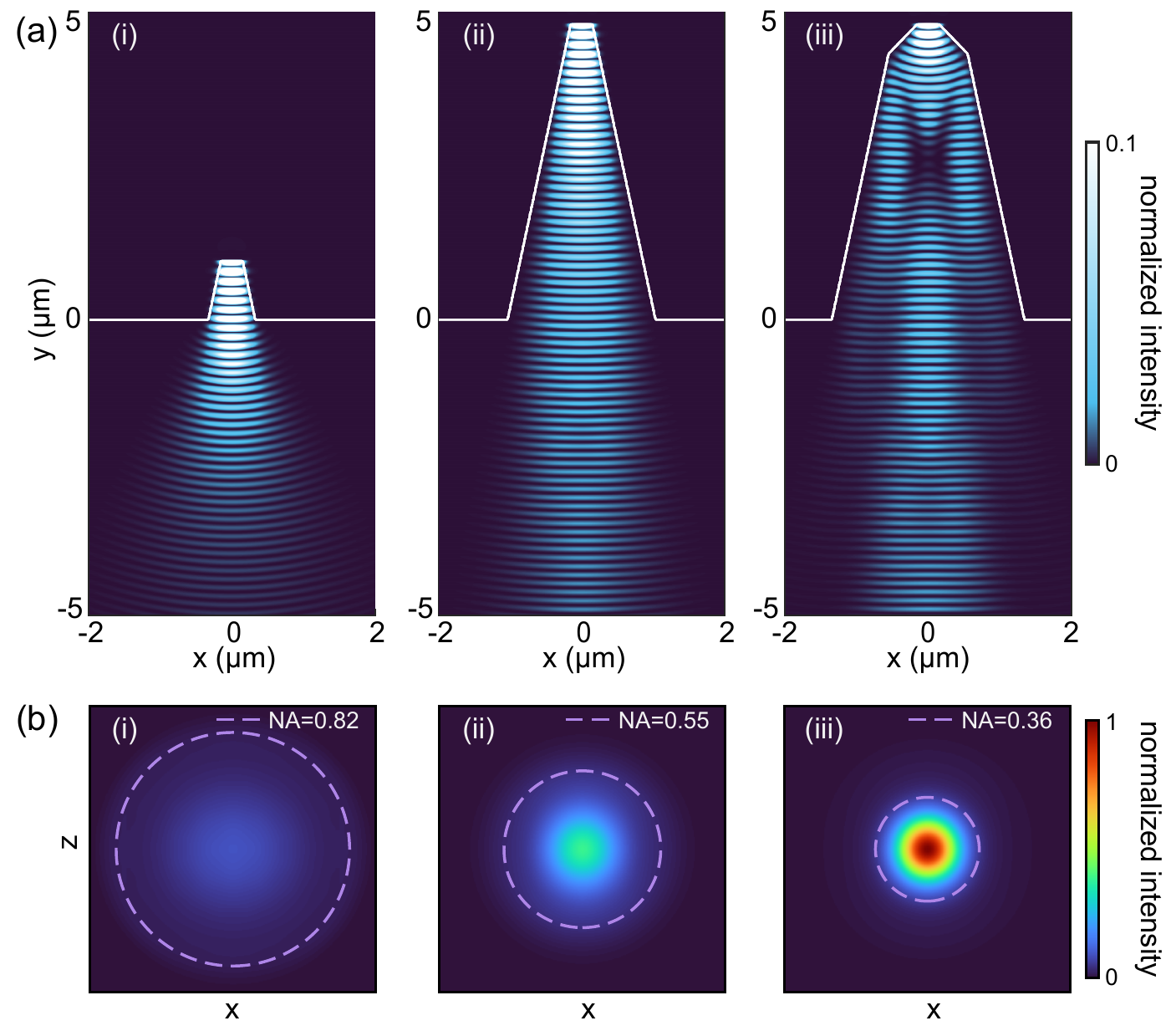}
   \caption{
   Illustrative mode simulations. (a) Propagating optical intensity for a fundamental-mode source (700 nm) in the diamond at the pillar facet ($R_\mathrm{{top}}=150$ nm) for a (i) 1-$\upmu$m-tall single cone, (ii) 5-$\upmu$m-tall single cone, and (iii) 5-$\upmu$m-tall multicone. Colorbars are saturated at 0.1 of the maximum intensity for each device to increase visibility. (b) Far-field intensity resulting from NV-center emission for the geometries in (a) normalized to the maximum intensity of the MC structure (b(iii)). Purple dashed lines indicate $\mathrm{NA}_{\,0.8}$.
   \label{fig:2}}
\end{figure}

To verify the design principles obtained from Fig. \ref{fig:1}, we simulate the propagating intensity of a monochromatic (700 nm), fundamental-mode excitation at the pillar facet for three representative structures (Fig. \ref{fig:2}a), including two SCs of different heights and one MC. As expected, the SCs in Figs. \ref{fig:2}a(i)-a(ii) exhibit an adiabatic expansion of the fundamental mode as it travels down the structure.~\cite{milton_mode_1977,fu_efficient_2014} For the 1-$\upmu$m-tall device, $\mathrm{\Delta}\approx2.2$; in contrast, the 5-$\upmu$m-tall pillar exhibits $\mathrm{\Delta}\approx6.9$ and superior collimation. In addition, the tall MC structure ($\phi=51^\circ$ and $H_1=4.5\ \upmu$m, Fig. \ref{fig:2}a(iii)) facilitates rapid modal expansion ($\mathrm{\Delta}\approx9.0$) as well as transverse-mode conversion.~\cite{milton_mode_1977,fu_efficient_2014} While the reduction in divergence angle depends on the exact modal composition, the exiting beam clearly exhibits flatter wavefronts than an SC of the same height (Fig.~\ref{fig:2}a(ii)). 

Finally, we simulate the far-field intensities for NV emission within the same geometries (Fig.~\ref{fig:2}b, see SI for wavelength-dependent plots, Fig. S1). Here, we normalize all plots to the maximum value obtained for the MC and overlay circles corresponding to $\mathrm{NA}_{\,0.8}$. As expected, superior collimation is obtained for the tall MC structure. Moreover, despite the presence of higher-order transverse modes, the far-field emission remains approximately Gaussian and is therefore compatible with fiber-coupled applications. 

\section{Fabrication}

\label{sec:fab}
\begin{figure}
\includegraphics{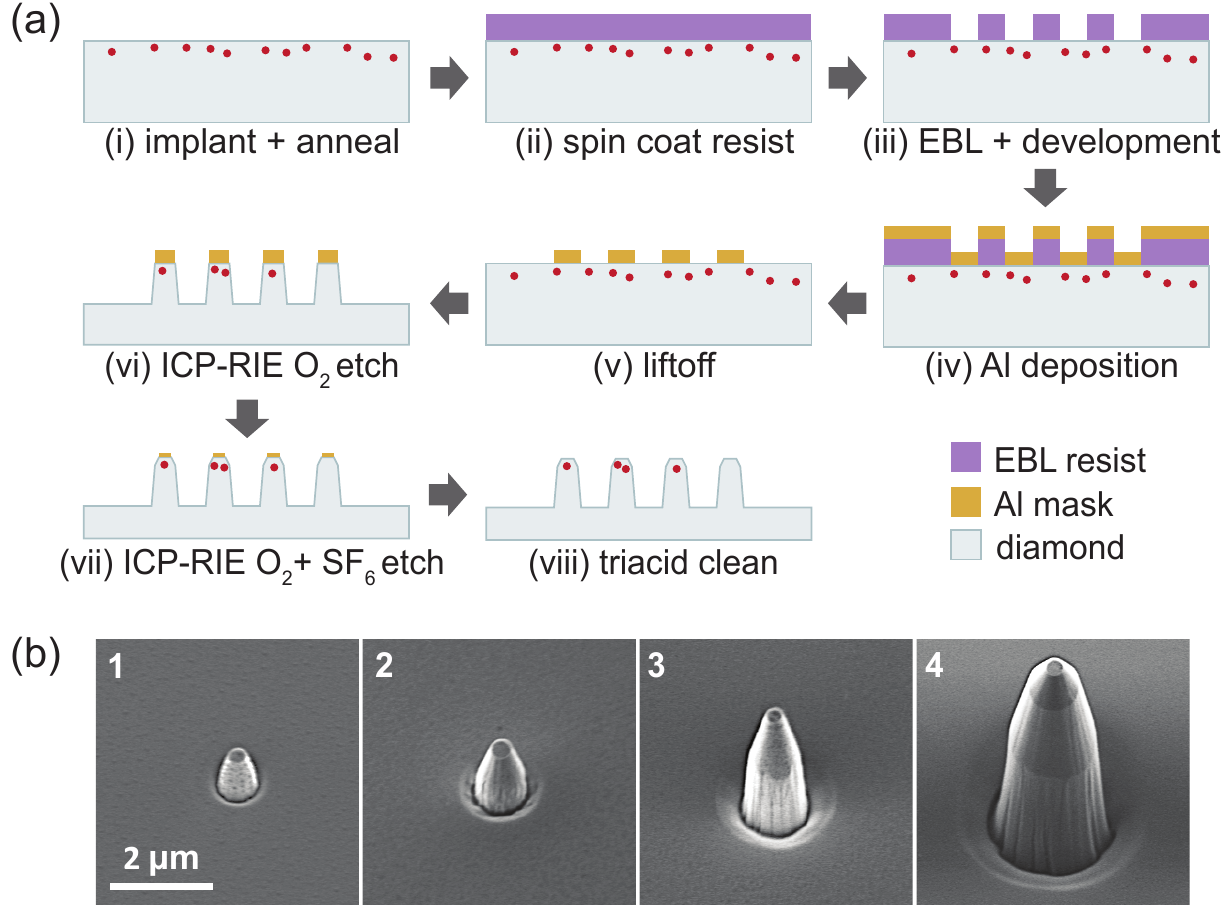}
  \caption{
  Nanofabrication process. (a) Pillar fabrication process flow (details in the main text and SI). (b) Scanning electron microscope (SEM) images of fabricated pillar geometries. Pillars 1-3 were etched using an O$_2$ chemistry, while pillar 4 was processed with O$_2$ plasma followed by a mixture of O$_2$ and SF$_6$.
\label{fig:3}}
\end{figure}

\noindent We verify our simulation results by fabricating four representative pillar geometries for experimental characterization (see SI for extended fabrication details). To ensure consistent material properties, all devices are created from the same diamond crystal that is sliced into 20-$\upmu$m-thick membranes (facilitating optical collection through the substrate). Figure~\ref{fig:3}a illustrates the fabrication process flow: first, a high-quality diamond surface is prepared by removing 6 $\upmu$m of material using an inductively coupled plasma reactive-ion etching (ICP RIE) process, which relieves crystal strain from polishing and smooths the surface to $<0.2$ nm-rms. The substrate is then cleaned in a boiling triacid mixture (1:1:1 ratio of HNO$_3$, H$_2$SO$_4$, and HClO$_4$) and implanted with $^{15}$N ions. Finally, implanted nitrogen is converted to NV centers {\it via} high-vacuum annealing. 

Next, we lithographically define four pillar geometries (pillars 1-4, Fig.~\ref{fig:3}b) using a layered electron-beam resist followed by electron-beam lithography. Specifically, we pattern circles with radii of 200, 275, 400, and 650 nm (for pillars 1-4, respectively). Following development, aluminum masks are deposited by electron-beam evaporation and transferred into the diamond by ICP RIE (details below). Etching is terminated when $R_{\mathrm{top}}\approx150$ nm, which is determined using a scanning electron microscope (SEM); consequently, the height of the final structure is defined by the initial mask diameter. After etching, the remaining aluminum mask is removed in a triacid mixture. 

Pillars 1-3 are created using an O$_2$-plasma recipe that is intermittently interrupted with SF$_6$ plasma to avoid micromasking. Initially, this process reduces the mask height while the diameter is largely unaffected, resulting in SC structures with $\theta= 78-86^{\circ}$. This operation modality is used to fabricate pillar geometries 1 and 2 (Fig.~\ref{fig:3}b(i)-(ii)), with $H=1.3$ $\upmu$m and $H=1.8$ $\upmu$m, respectively. However, prolonged plasma exposure (required for tall pillars) eventually causes lateral mask erosion, which reduces the sidewall angle and forms MC structures.~\cite{forsbergInclinedSurfacesDiamond2013} Indeed, pillar 3 is fabricated using the same plasma recipe but exhibits a second, shallow sidewall angle with $\phi=78^\circ$ (Fig.~\ref{fig:3}b(iii), $H_1=2.1\ \upmu$m and $H_2=1.9\ \upmu$m). 

While promising, fabrication of MC structures by lateral mask erosion is difficult to control and requires constant monitoring with SEM. Consequently, we develop an alternative method for achieving a second, shallow sidewall angle based on a new plasma chemistry (4:1 flow ratio of O$_2$ and SF$_6$, respectively) with different etch selectivity.
Subsequently, we create the MC geometry in Fig. \ref{fig:3}b(iv) by first etching with the O$_2$ recipe ($H_1=5.2\ \upmu$m) followed by the new plasma mixture ($H_2=0.9\ \upmu$m). Interestingly, the resulting structure exhibits two additional sidewall angles: $\phi=64^\circ$ ($H=5.2-6.1\ \upmu$m) from the new chemistry, and $\phi=82^\circ$ ($H=2.9-5.2\ \upmu$m) caused by mask erosion during the O$_2$ process. Simulations of this structure suggest that the additional tapering at the facet further improves directionality and collimation of the output beam. 

\section{Characterization}

\begin{figure}
\includegraphics{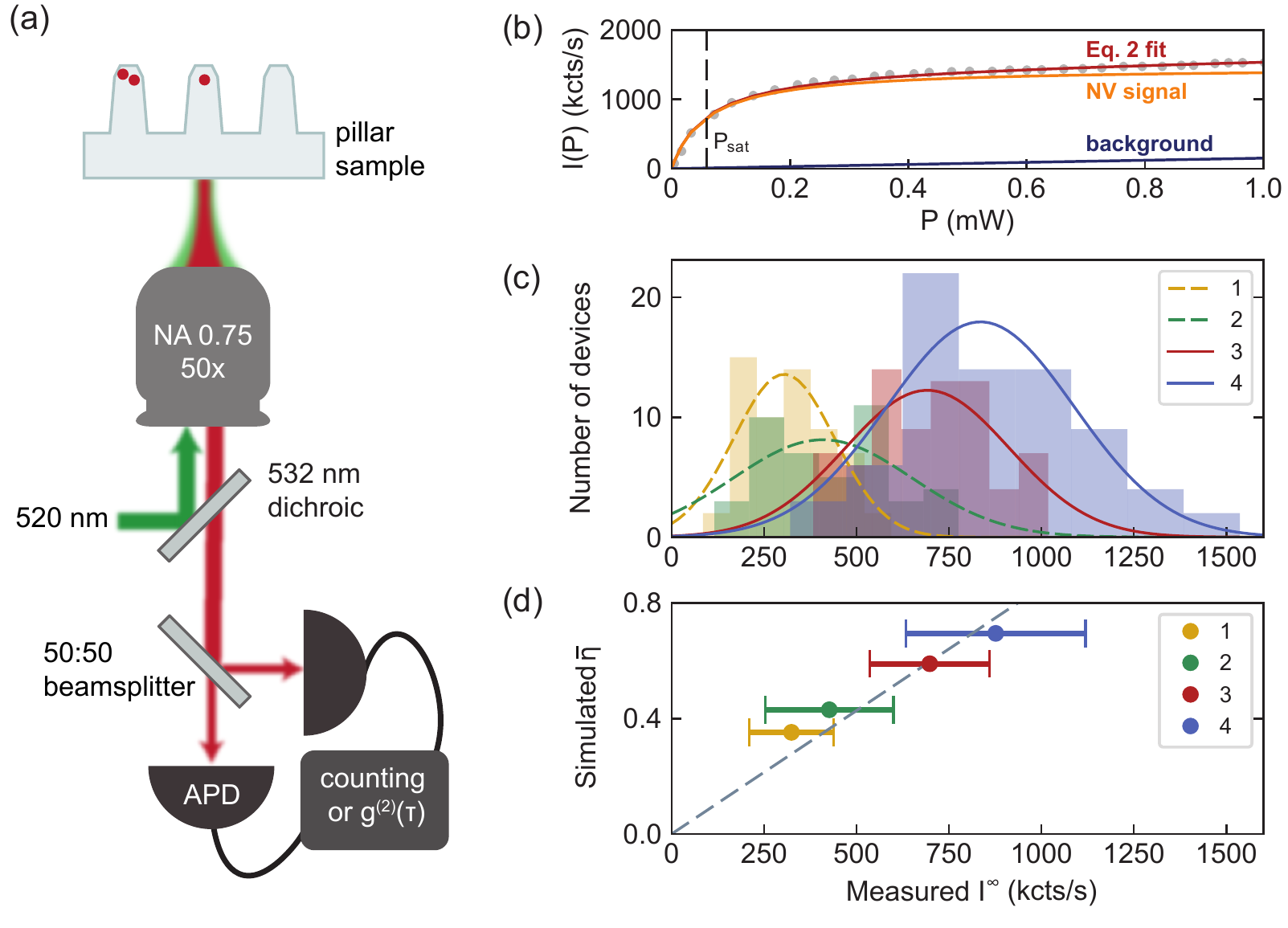}
\caption{\label{fig:4}
Experimental characterization. (a) A simplified schematic of the confocal microscope. (b) An example fluorescence saturation measurement for a multicone device (pillar 4) fit with Eq.~\ref{eqn:saturation} ($P_{\mathrm{sat}}=59\pm 2$ $\upmu$W, $I^\infty=1464.9\pm9.8$ kcts/s). 
(c) Histogram showing the number of viable devices {\it vs.} $I^\infty$ for the four geometries with Gaussian fits. (d) Simulated $\bar{\eta}$ {\it vs.} measured $I^{\infty}$ for each geometry with a linear fit overlaid. 
}
\end{figure}

Nanopillars are characterized using the confocal microscope illustrated in Fig. \ref{fig:4}a. We excite NV centers using a home-built 520-nm excitation laser reflected at a dichroic beamsplitter (Semrock SEM-FF526). Subsequently, the beam is scanned using galvo mirrors onto the back of an objective lens (Mitutoyo Plan Apo HR $50\times$, $\mathrm{NA}=0.75$) and is focused onto the sample. Fluorescence is collected using the same objective lens, transmitted through the dichroic, spectrally filtered (Semrock SEM-FF01-709/167), split by a 50:50 beamsplitter, and coupled into two optical fibers (SMF28). These fibers exhibit multimode operation over the NV emission spectrum; consequently, we assume that all light within the experimental NA is collected. Finally, collected photons are sent to fiber-coupled avalanche photodiodes (APDs, Laser Components COUNT\textsuperscript{\textregistered}) and the recorded counts are either summed to obtain the total NV fluorescence or used as inputs for a time-correlated single-photon-counting system (TCSPC, PicoQuant Picoharp 300). 

As a first test, we identify devices containing single NVs from intensity autocorrelation measurements~\cite{brouri_optlett_2000} taken with the TCSPC (see SI for details). We survey over 400 pillars of each geometry, yielding 59, 43, 73, and 74 viable structures for pillar geometries 1-4, respectively. 

Next, we compare the performance of each structure by measuring the NV fluorescence at infinite pump power ($I^\infty$), which scales linearly with $\bar{\eta}$. $I^\infty$ is obtained by fitting power-dependent fluorescence measurements (Fig. \ref{fig:4}b) to
 \begin{align}
I(P)=I^\infty \frac{P}{P+P_{\mathrm{sat}}}+c_{\mathrm{bg}}P,
\label{eqn:saturation}
\end{align}
where $I$ is the measured count rate, $P$ is the excitation power, $P_{\mathrm{sat}}$ is the fitted saturation
power, and $c_{\mathrm{bg}}$ is a linear background contribution. The results are illustrated in a histogram (Fig.~\ref{fig:4}c) showing number of devices {\it vs.} $I^{\infty}$ for the four geometries. Indeed, the increase in collection intensity afforded by tall pillars and the MC geometry confirms our simulation predictions.
A quantitative comparison can be gained by plotting simulated $\bar{\eta}$ against measured $I^{\infty}$ for each test geometry (Fig.~\ref{fig:4}d), yielding a linear relationship with slope $(850\pm50)\times10^{-5}$ s/kct. 

\section{Discussion}

\begin{figure}  \includegraphics{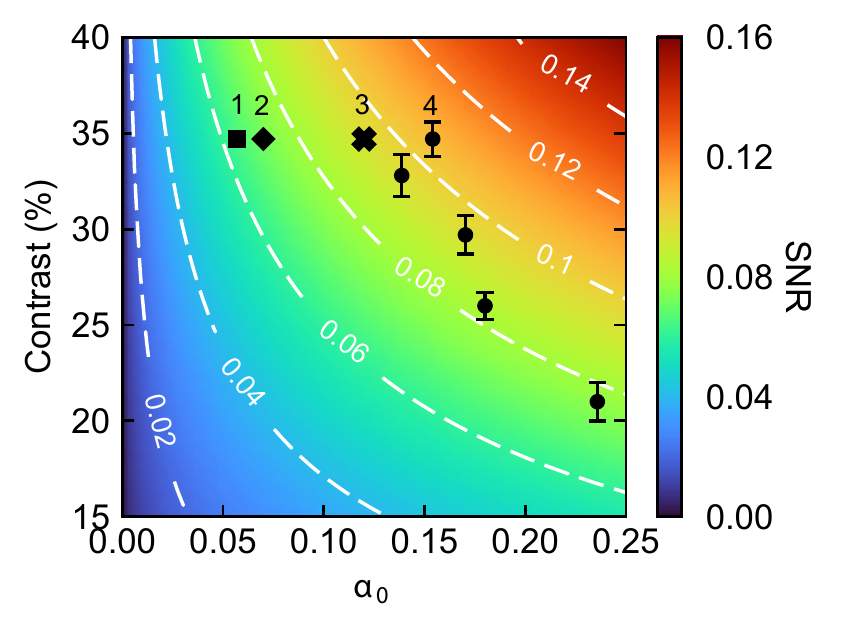}
  \caption{
  Measured optical contrast ($C$) {\it vs.} integrated photon counts per measurement ($\alpha_0$) for a representative pillar 4 device at five different excitation powers (circular markers). The corresponding spin-readout SNR is calculated and overlaid in a color plot. The maximum experimental $\mathrm{SNR}=0.106$ is obtained for $\alpha_0=0.154$ and $C=34.7\pm0.9\%$. For comparison, we estimate the corresponding maximum SNR for pillars 1-3 as 0.064, 0.073, and 0.095, respectively.
  \label{fig:5}}
\end{figure}

Finally, we explore how the improvement in collection efficiency afforded by an optimized pillar structure impacts sensor performance. We select the spin-readout SNR for a single measurement as our figure of merit; the magnetic sensitivity (or minimum detectable field) scales inversely with this quantity.~\cite{janitz2022diamond}

We consider an NV sensing experiment where the measured signal is encoded in the populations of two ground-state spin sublevels, denoted $\ket{0}$ and $\ket{1}$. These populations can be discriminated by their integrated spin-dependent fluorescence, $\alpha_{0}$ and $\alpha_{1}$, due to the optical contrast between spin states ($C=1-\alpha_1/\alpha_0$). Typically, $\alpha_{0/1}\ll 1$ for off-resonant optical readout, and measurements are dominated by shot noise, yielding~\cite{hopperSpinReadoutTechniques2018}

\begin{equation}   \mathrm{SNR}\approx\sqrt{\alpha_0}\frac{C}{\sqrt{2-C}}.
\label{eqn:SNR}
\end{equation}
 
We estimate this quantity for a representative pillar 4 device by measuring $\alpha_0$ and $C$ at different laser powers (circular markers in Fig.~\ref{fig:5}, see SI for details). Here, we apply a small magnetic field (B $\approx$ 2 mT) along the NV axis to lift the degeneracy of the $m_{s} = \pm1$ ground states and work within the $m_{s} = -1$ ($\ket{1}$) and $m_{s} = 0$ ($\ket{0}$) manifold. We estimate $C$ at each laser power by fitting power-dependent Rabi experiments and calculate the average $\alpha_0$ per measurement by integrating the spin-dependent fluorescence for a fixed measurement time $T_\mathrm{m}=300$ ns. While we use a fixed value for simplicity, $T_\mathrm{m}$ can be optimized at each laser power to maximize SNR.~\cite{gupta2016efficient} Subsequently, we overlay the calculated values from Eq.~\ref{eqn:SNR} for comparison, yielding a maximum $\mathrm{SNR}=0.106$ for $C=34.7\pm0.9\%$ and $\alpha_0=0.154$ photons/measurement. 
Consequently, $\mathrm{SNR}=1$ could be achieved {\it via} integration of only 90 measurements. 

For comparison, we estimate the corresponding peak SNR for pillars 1-3 (square, diamond, and cross markers in Fig.~\ref{fig:5}, respectively). Here, we assume the same optical contrast of $C=34.7\pm0.9\%$ since the internal NV dynamics and background fluorescence should not change between device geometries.~\cite{momenzadehNanoengineeredDiamondWaveguide2015} Moreover, we scale $\alpha_0=0.154$ by the ratio of average $I^\infty$ values obtained for each structure (Fig.~\ref{fig:4}d), resulting in a spin-readout SNR of 0.064, 0.073, and 0.095, for pillars 1-3, respectively. Based on these results, pillar 4 should exhibit a factor-of-three reduction in measurement time to obtain the same SNR as pillar 1. 
   
\section{Conclusion}
In this work, we present an optimized diamond nanopillar structure with enhanced optical collection efficiency. 
First, we develop a simulation model that predicts superior performance for tall ($H\geq5\ \mu$m), conical structures with an additional taper near the facet. Next, we fabricate four test geometries using a novel two-step etching process; subsequent optical characterization verifies our design principles, yielding an experimental spin-readout $\mathrm{SNR}=0.106$ for an optimized device. 

Moving forward, the improved collection efficiency afforded by the optimized MC device will greatly benefit a number of emerging quantum-sensing technologies. First, the integration time required to achieve a given SNR scales inversely with $\bar{\eta}$, yielding experimental speed-up. Moreover, crucial for scanning applications, the MC allows for a reduction in $R_{\mathrm{top}}$ compared to the SC design. Consequently, such structures will simultaneously enable exquisite topographic resolution and excellent measurement sensitivity. Lastly, the resulting beam of the optimized MC device is highly Gaussian and collimated, and therefore compatible with free-space or fiber-coupled technologies utilizing low-NA collection optics. 

Beyond NV centers, the devices developed here could be easily translated to alternative diamond defects, including the negatively charged group-IV emitters~\cite{bradac2019quantum} and other emerging color centers\cite{rose2018observation,telecom_oband_diamond}. Finally, our design principles are also applicable to emitters in alternative materials, such as defect centers in SiC~\cite{lukin2020integrated} or rare-earth ions in doped glasses.~\cite{zhong2018optically}

\begin{acknowledgement}
The authors thank Stefan Ernst, Zhewen Xu, Marius Palm, Pol Welter, Mihai Gabureac, Simon Josephy, Andrea Morales, Laura Alicia V\"{o}lker, and John Abendroth for fruitful discussions. This work was supported by the European Research Council through ERC CoG 817720 (IMAGINE), the Swiss National Science Foundation (SNSF) through Project Grant No. 200020-175600 and NCCR QSIT, a National Centre of Competence in Research in Quantum Science and Technology, Grant No. 51NF40-185902, and the Advancing Science and TEchnology thRough dIamond Quantum Sensing (ASTERIQS) program, Grant No. 820394, of the European Commission. EJ acknowledges support from  a Natural Sciences and Engineering Research Council of Canada (NSERC) postdoctoral fellowship (PDF-558200-2021).
\end{acknowledgement}

\begin{suppinfo}
The SI provides additional information on wavelength-dependent simulations, fabrication details, single-photon measurements, NV-displacement simulations, and Rabi-oscillation measurements. 
\end{suppinfo}

\bibliography{multicone_main_arxiv}

\end{document}